\documentclass[]{proceed}
\usepackage{epsfig} 
\usepackage{times} 
\usepackage{citesort} 

\begin{document}

\title{A minimal model for slow dynamics: Compaction of granular
       media under vibration or shear}

\titlerunning{A minimal model for compaction}

\author{Stefan Luding \inst{1}
        \and Maxime Nicolas \inst{2}
        \and Olivier Pouliquen \inst{2}
        }


\institute{Institute for Computer Applications 1, \\
Pfaffenwaldring 27, 70569 Stuttgart, Germany \\
e-mail: lui@ica1.uni-stuttgart.de\\
\and Groupe Ecoulement de Particules, IUSTI, Technopole de Chateau-Gombert, \\
     5 rue Enrico Fermi, 13453 Marseille Cedex 13, France
      }
\maketitle

\begin{abstract}
Based on experiments of the compaction of granular materials
under periodic shear of a packing of glass beads, a minimal model 
for the dynamics of the packing density as a function of time is 
proposed.  First, a random ``energy landscape'' is created
by a random walk (RW) in energy. Second, an ensemble of RWs
is performed for various temperatures in different temperature-time
sequences. We identify the minimum (mean) of the energy landscape with 
the maximum (random) density. 
The temperature scaled by the step-size of the energy landscape
determines the dynamics of the system and can be regarded as the 
agitation or shear amplitude. The model reproduces qualitatively 
both tapping and shear experiments.
\vspace{0.1cm}~
\end{abstract}

\section{Introduction}

The issue of slow dynamics, rare events and anomalous diffusion 
is object to ongoing research in statistical physics 
\cite{bouchaud90,schiessel97,rajasekar98,laloux98,fisher98}.
One example for an experimental realization of slow dynamics is 
the compaction of granular materials, being of interest for both
industrial applications and research.  Many compaction experiments 
have been carried out in the last decades,
see e.g.~\cite{ayer66,kuno82,aydin97,ewsuk97}, but the 
evolution of density with time is far from being well understood.
Recent laboratory experiments concern pipes filled
with granular media and periodically accelerated in
order to allow for some reorganization
\cite{knight95,nowak97,nowak98}.  The compaction
dynamics was obtained to be logarithmically slow and could
be reproduced with a simple parking model \cite{ben-naim96,kolan99}.  
Alternative experiments concern
a sheared block of a granular model material (monodisperse
glass spheres) and display a similar dynamics \cite{nicolas99}.
Numerical model approaches, like a 
frustrated lattice gas (the so-called ``Tetris'' model)
\cite{nicodemi95,caglioti97,piccioni99}
also lead to this slow dynamic behavior, as well as
some theory based on stochastic dynamics \cite{linz99}.
The fact that a peculiar dynamics is reproduced by so many
models indicates that it is a basic and essential phenomenon.

Rather than modeling granular systems in all details, e.g.~in the 
framework of molecular dynamics or lattice gas simulations
\cite{peng97b,piccioni99}, we will propose a very simple model based
on the picture of a random walk in a random energy landscape,
for a review see \cite{bouchaud90}, and even simpler than recent,
very detailed considerations in the same spirit \cite{head98}.
Random walks have been examined, for example, on fractals and ultrametric 
spaces, where the continuous time random walk was introduced in order
to allow a mathematical treatment \cite{kohler91}.
If a random walker is situated in a uncorrelated, random, fractal energy 
landscape, the process is called Sinai-diffusion \cite{bouchaud90}.
The aim of this study is to show that the Sinai model is in qualitative
agreement with the compaction dynamics of granular media.  
An issue not addressed here is a quantitative adjustment which
can be reached, for example, by introducing correlations in
the EL.

\section{Summary of the experimental results}

The subject of compaction of granular material 
\cite{ayer66,kuno82,aydin97,ewsuk97} has been 
recently revisited through careful experiments carried out by Knight 
et al.~and Nowak et al.~\cite{knight95,nowak97,nowak98}.
The experiment consists of a vertical cylinder, full of monodisperse 
beads, which is submitted to successive distinct taps of 
controlled acceleration (vertical vibration).  The measurement of the 
mean volume fraction after each tap gives a precise information about 
the evolution of the compaction.  From the first experiments performed
with taps of constant amplitude, the increase in volume fraction was found 
to be a very slow process well fitted by the inverse of a logarithm of the 
number of taps.  Nowak et al.~\cite{nowak97,nowak98} have then 
studied the compaction under taps of variable 
amplitude, and showed that irreversible processes occur during the 
compaction.  Starting with a loose packing, the evolution of the 
volume fraction is not the same when increasing the amplitude of 
vibration as when decreasing. The first branch appears to be irreversible, 
whereas the second is reversible.

\begin{figure}[tb]
\begin{center}
\epsfig{file=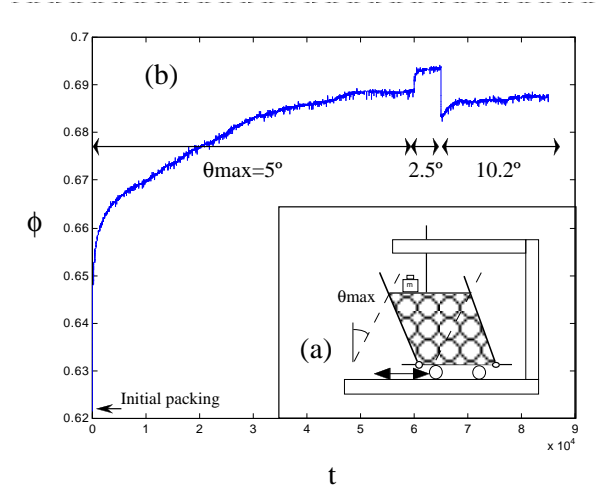,height=6.5cm,angle=0}
\end{center}
\caption{(a) Sketch of the experimental setup. (b) Evolution of the
volume fraction as function of shear cycles $n$: a saturated state is 
seemingly reached after $6\times10^4$ cycles 
with an initial shear angle of 5 degrees; a positive jump in volume 
fraction is observed when the angle is decreased to 2.5 degrees and 
applied for $5\times10^3$ cycles; a negative jump is observed when the 
angle is increased to 10.2 degrees.
}
\label{fig:exp}
\end{figure}

We have recently performed a compaction experiment based on cyclic 
shear applied to an initially loose granular packing \cite{nicolas99}.  
A parallelepipedic box full of beads is submitted to a horizontal shear 
through the periodic motion of two parallel walls at amplitude 
$\theta_{\rm max}$, see Fig.\ \ref{fig:exp}(a). Compaction occurs during this 
process, leading to crystallisation of the beads in the case of a monodisperse 
material. The control parameter in this configuration is the maximum 
amplitude of shear $\theta_{\rm max}$ (inclination angle of the walls). 
The measurement of the mean volume fraction $\phi$ shows that 
compaction under cyclic shear is a very slow process as in the 
vertical vibration experiments (typically $5\times10^4$ shear cycles or 
taps). The higher the shear amplitude the more efficient is the 
compaction (shear amplitudes up to $\theta_{\rm max} < 12.5^{\circ}$ were
examined), when starting with a loose packing of the same initial 
volume fraction. More surprising results arise when the packing is 
submitted to a sudden change in shear amplitude.  We have observed 
that a ``jump'' in volume fraction occurs which is opposite and 
proportional to the change in $\theta_{\rm max}$.  Sudden increase 
(resp.  decrease) in the shear amplitude decreases (resp. increases) 
the volume fraction (Fig.\ \ref{fig:exp}). The response is very rapid (less 
than 20 cycles) and quasi-independent of the state before the angle 
change. For more detailed experimental results see \cite{nicolas99}. 
~
\section{The Model}
A naive picture that evolves out from the experimental observations is the 
analogy between the packing of beads and a thermal system seeking for 
a minimum of energy in a very complex potential-energy landscape 
\cite{bouchaud90}.  
Due to some agitation (shear)
compaction occurs and the total potential energy of the packing 
decreases. Starting from a loose packing at $n=0$, the vibrational 
or shear excitation can be seen as the analog of the temperature in the sense 
that the excitation allows for an exploration of the phase space. In a 
granular packing of monodisperse spheres, the absolute minimum of the energy 
is obtained for a perfect (fcc) or (hc) crystal with volume fraction 
$\phi=0.74$. If the energy landscape is complex with a lot of different scale 
valleys or hills, one can understand that an efficient, fast compaction will 
be obtained with high temperatures: the system is then able to escape the 
deep local valleys and find a valley with lower potential energy.  
A decrease of the temperature is then needed to explore local fine-scale 
minima.  The goal of this paper is simply 
to explore this idea by studying the dynamics of random walkers on a 
random landscape (the Sinai model), and to
show that this very naive picture gives results in 
qualitative agreement with the experimental observations. 

\begin{figure}[htb]
\begin{center}
\epsfig{file=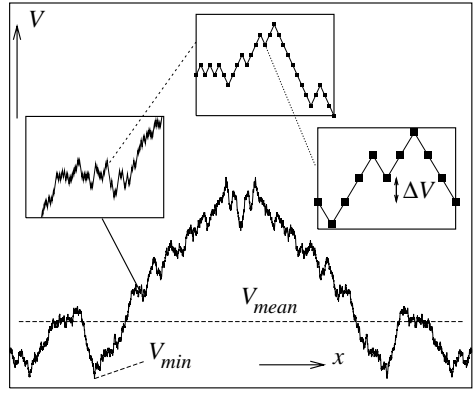,height=6.5cm,angle=-90}
\end{center}
\caption{Schematic plot of a typical energy landscape as used 
for the simulations. The insets show different zoom levels, in order to get 
an idea of the local situation. For an explanation of the symbols,
see the text.
}
\label{fig:EL}
\end{figure}

Our model is based on the assumption that all possible
configurations of a granulate in a given geometry
can be mapped onto an ``energy landscape'' (EL).  Since a 
simple two-level system (as used to model the dynamics in simple glasses) 
does not lead to the experimentally observed phenomenology,
we assume a fractal energy landscape created
by a random walk in energy with phase space coordinate
$x$.  A typical EL with energy $V(x)$ is schematically shown 
in Fig.\ \ref{fig:EL}.
The stepsize in energy is $\Delta V$, the mean of the 
energy landscape is $V_{\rm mean}$ and its absolute minimum 
is $V_{\rm min}$.  Here, the EL is symmetric to its 
center, in order to allow for periodic boundary 
conditions in $x$. 
Given some EL, the granulate is now modeled as an ensemble of random 
walkers diffusing on the EL, with a temperature $T_{RW}$. 
The analog to the density of a granular packing is the
rescaled energy of an ensemble of random walkers in the energy landscape
\begin{equation}
\nu = 1 - \frac{E-V_{\rm min}}{V_{\rm mean}-V_{\rm min}} ~,
\end{equation}
which we denote as {\em density} in the following.
For the (random) initial configuration, the energy will be 
$E \approx V_{\rm mean}$ so that $\nu=0$; for the close-packing
configuration, i.e.~all RWs are in the absolute 
minimum, one has $E \approx V_{\rm min}$ and thus $\nu=1$.
The energy landscape has the
size $L$, the ensemble of RWs consists of $R$ random walkers
and $S$ is the number of steps performed.

\begin{figure}[htb]
\begin{center}
\epsfig{file=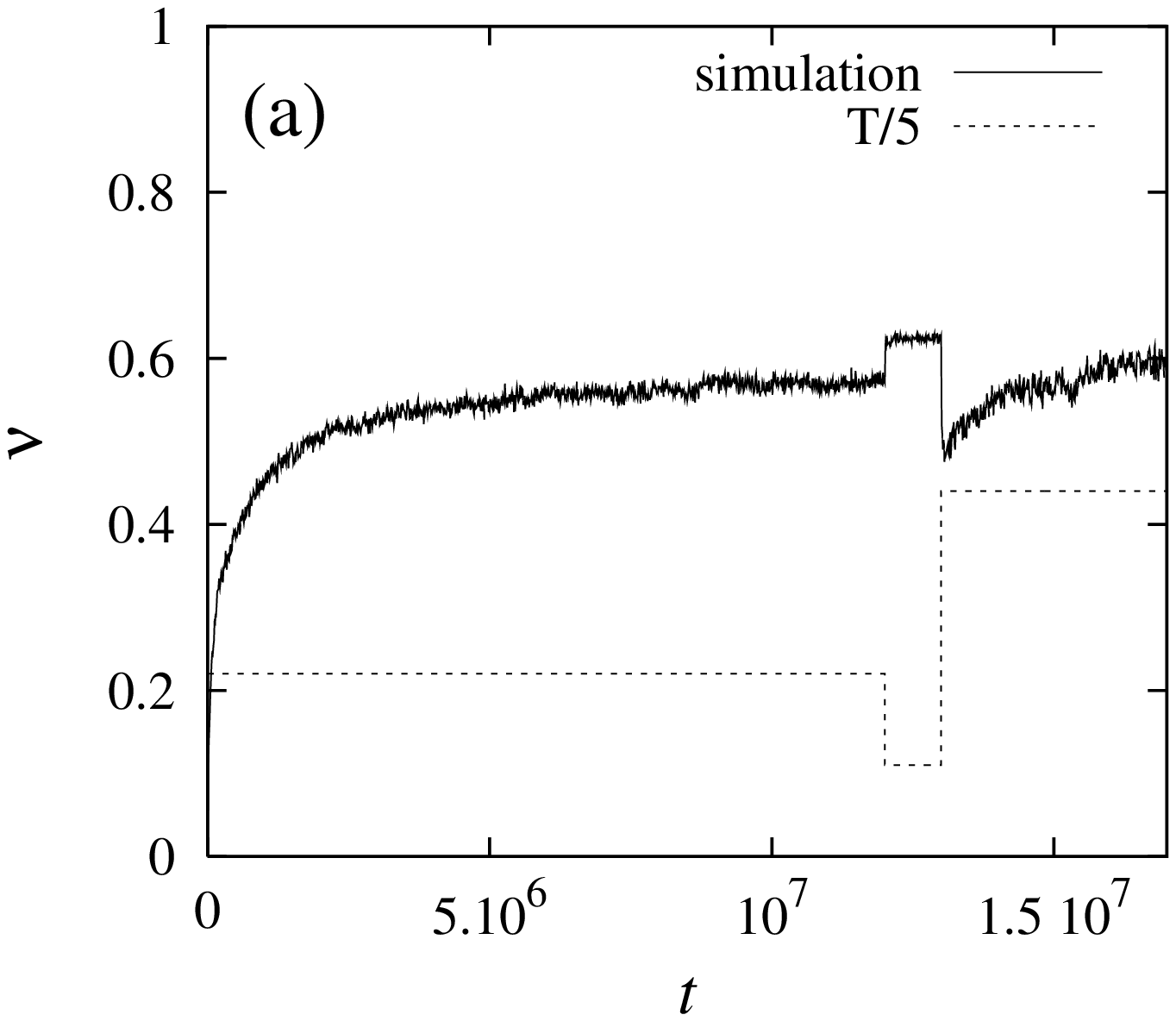,height=6.0cm,angle=-90}
\epsfig{file=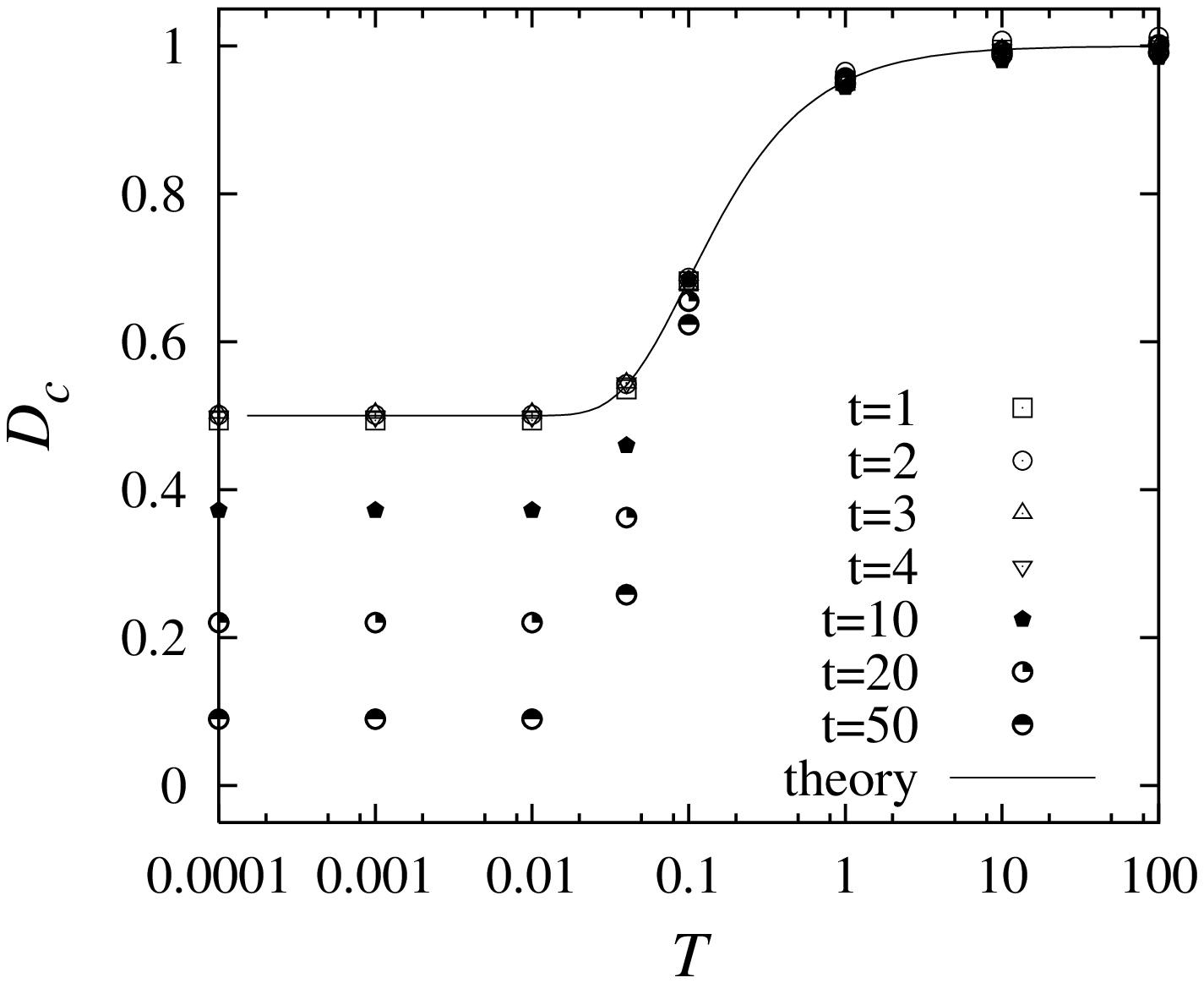,height=6.0cm,angle=-90}
\end{center}
\caption{(a) Density plotted against time $t$ (in units of 
Monte-Carlo steps) for a simulation with $T_0=1.1$, $L=5000$ and
$R=200$. The temperature-sequence is indicated by the dashed line.
(b) Diffusion constant $D_c=R_2(t)/\sqrt{t}$ 
at time $t$ as function of temperature $T$ for systems with
constant (initial) density. The solid line shows Eq.\ (\ref{eq:Dc}).
}
\label{fig:tav}
\end{figure}

Since the energy of the RWs in the EL corresponds to the potential energy 
of the packing, we interpret the (constant) stepsize $\Delta V$ (used here 
as one possibility for the construction of the EL) as a 
typical activation energy barrier. The maximum of $V$ corresponds to a 
random loose, local packing density, the mean to the (initial) random 
close packing, and the absolute minimum to the hexagonal close packing.
For one RW, the probability to jump within time $\Delta t$
from one site $x_i$ to its neighbor-site $x_{i \pm 1}$ is 
\begin{equation}
p_\pm(x_i) = \min \left [ 1, \exp(-\Delta_\pm(x_i)/T_{RW}) \right ] ~,
\end{equation}
with $\Delta_\pm(x_i)=V(x_{i \pm 1})-V(x_i)$. 
Since $\Delta V$ is the only energy scale of the system,
we define the dimensionless energy steps 
$\delta^i_\pm = \Delta_\pm(x_i)/\Delta V$ and the dimensionless
temperature $T=T_{RW}/\Delta V$.  Written in dimensionless
parameters, the jump probabilities are thus
$p^i_\pm = p_\pm(x_i) = \min \left [ 1, \exp(-\delta^i_\pm/T) \right ]$,
so that a particle always jumps downhill ($\delta^i_\pm \le 0$),
but jumps uphill only with a probability $e_0 = \exp(-1/T)$
(for $\delta^i_\pm > 0$), at finite temperature. The limits
$T \rightarrow 0$ and $T \rightarrow \infty$ correspond thus
to immobile particles or to a homogeneous RW, respectively.
The discrete master equation for the probability density $n^i(t)$, to 
find a particle at time $t$ at site $i$ of the EL, is
\begin{equation}
n^i(t+\Delta t)-n^i(t) =
 - \left [ p^i_+ + p^i_- \right ] n^i(t) 
 + p^{i+1}_- n^{i+1}(t) + p^{i-1}_+ n^{i-1}(t)
\label{eq:MEQ}
\end{equation}
and is (straightforwardly) simulated with $R=200$ random walkers
in an energy landscape with $L=5000$ sites, if not explicitly 
mentioned.  The time interval $\Delta t$ corresponds to one Monte-Carlo
step but {\em not} to one shear-cycle $n$. 
For the sake of simplicity, we measure $x$ in units
of the distance between neighboring sites $\Delta x = x_{i+1}-x_i=1$,
and time in units of $\Delta t$.  
The diffusion constant of a homogeneous random walk ($p_\pm=1/2$ or
$T \rightarrow \infty$) is thus $D=\Delta x^2/\Delta t=1$.  For a constant 
occupation probability (initial density $n_c=n^i(t=0)=R/L={\rm const.}$), one 
can extract the diffusion constant $D_c$ as function of the temperature, since
$p_\pm=1/2$ and $p_\pm=e_0/2$ occur with equal probability, so that
\begin{equation}
D_{c}(T) = \frac{1+e_0}{2} = \frac{1+\exp(-1/T)}{2} ~.
\label{eq:Dc}
\end{equation}
In Fig.\ \ref{fig:tav}, $D_c=R_2(t)/\sqrt{t}$ is plotted against $T$ 
after different times $t$, with $R_2(t)=\langle (x(t)-x_0)^2\rangle$. 
For large $T$, the system does not feel the EL and behaves like 
an ensemble of homogeneous RWs, whereas its behavior becomes subdiffusive 
after several steps for small $T$ when the EL is explored.
In a situation with $T \approx 0$, after a rather short transient, all RW
will occupy a local minimum, a valley ($\lor$), wheras the  
unstable local configurations hill ($\land$), or left- ($/$) and
right-slope ($\backslash$) cannot be occupied.
if $T$ is small enough, the random walkers stay trapped.
Note that this statement is strictly true for $T=0$ or when a fixed number 
of shear-cycles or taps is implied. In the Sinai diffusion model, for 
a finite system, the RW will always find the global minimum -- 
the temperature only determines the time-scale of this process 
\cite{bouchaud90}.  Since the duration of an experiment is 
limited, the global minimum cannot be found by all particles if the 
phase space volume $L$ is large enough.

\section{Results and Discussion}

\begin{figure}[htb]
\begin{center}
\epsfig{file=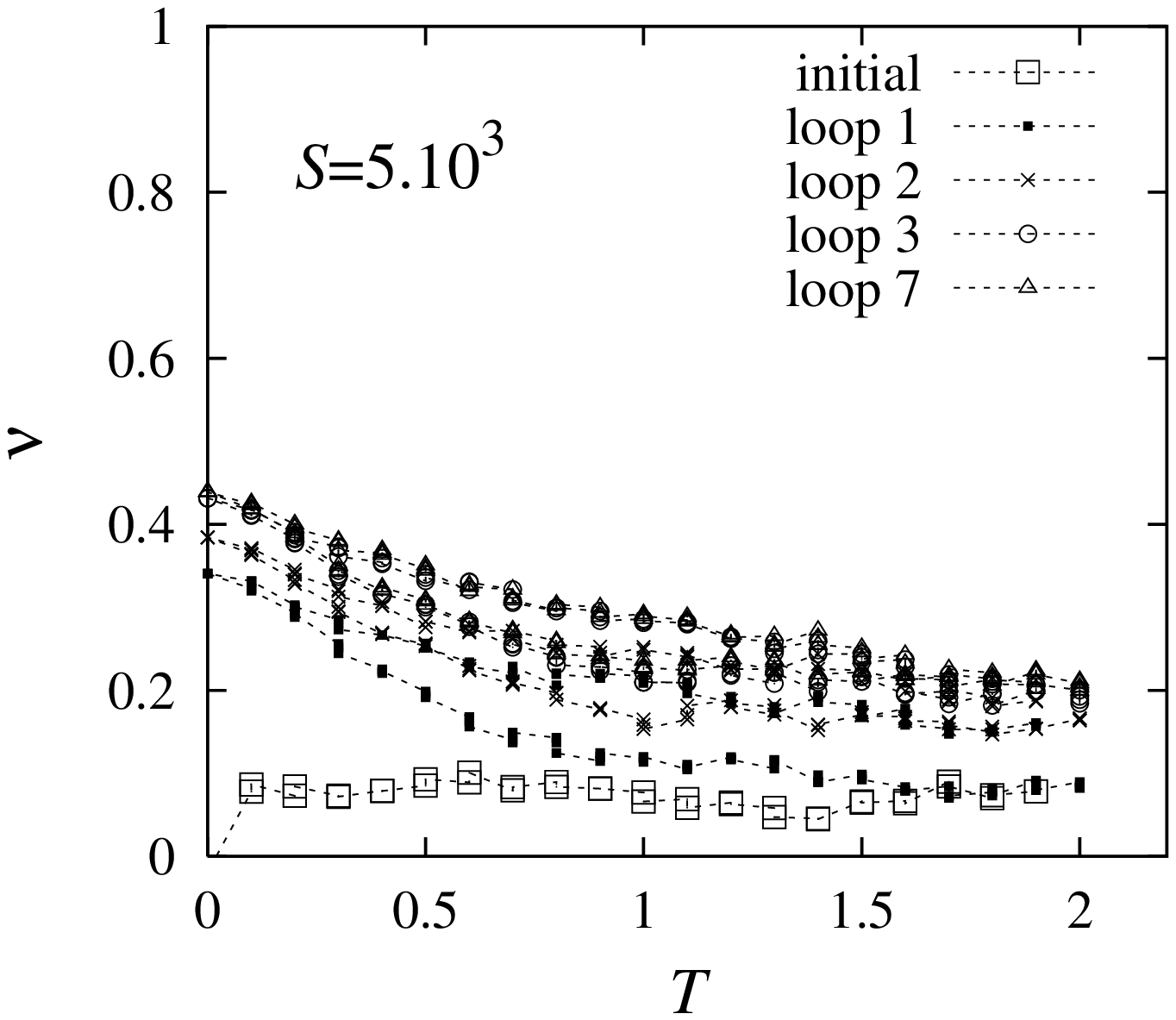,height=6.0cm,angle=-90}
\epsfig{file=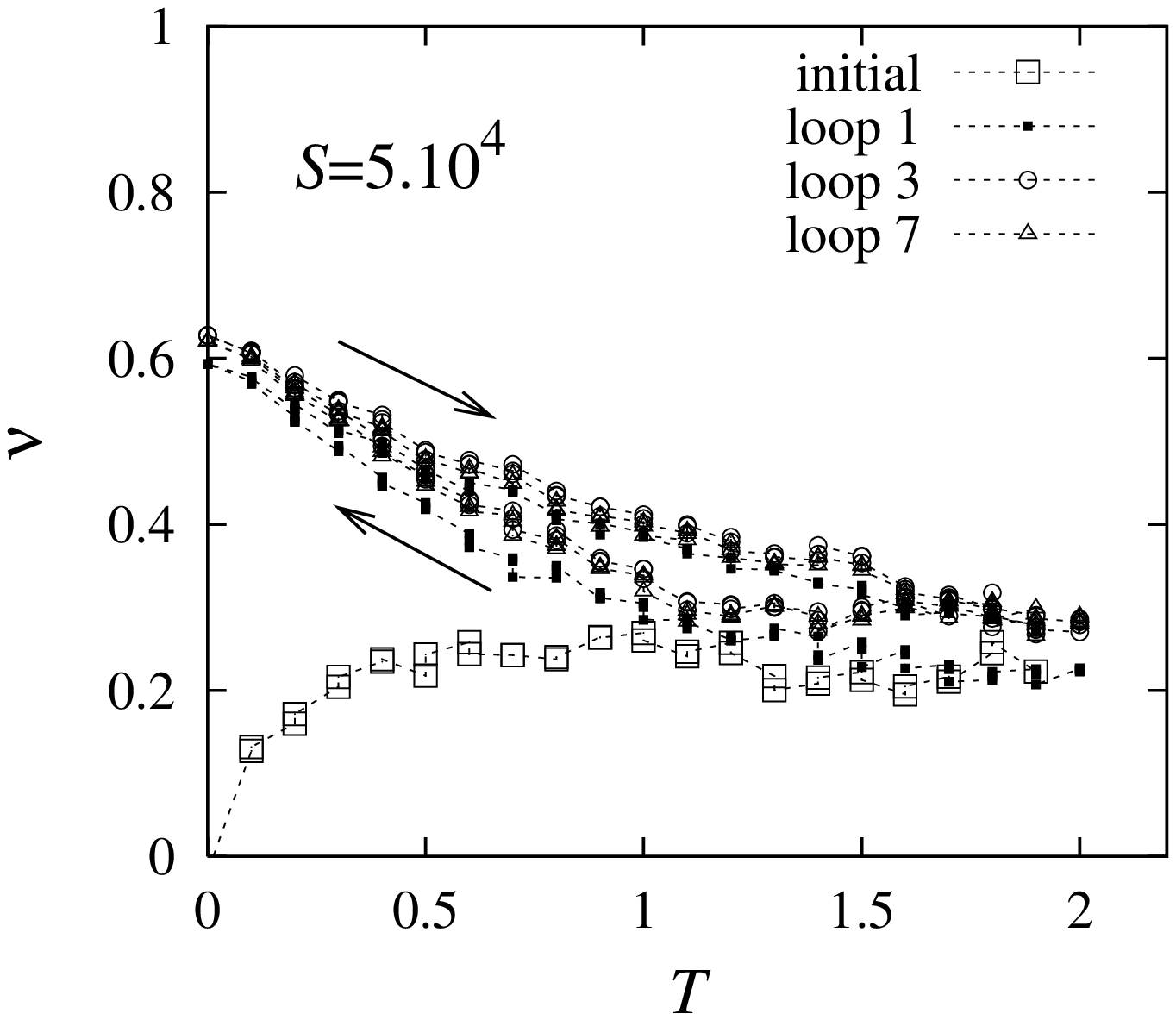,height=6.0cm,angle=-90}
\epsfig{file=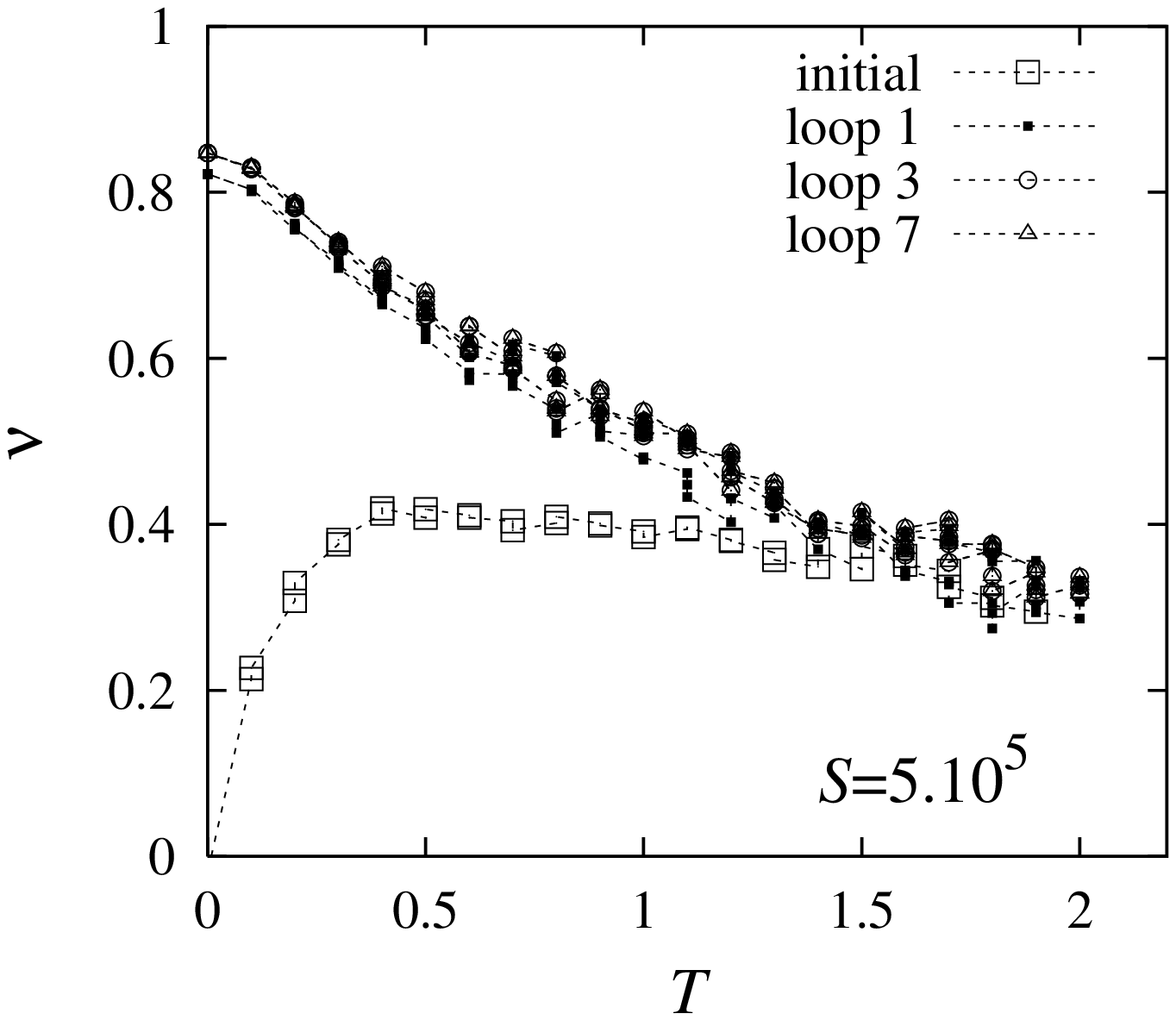,height=6.0cm,angle=-90}
\epsfig{file=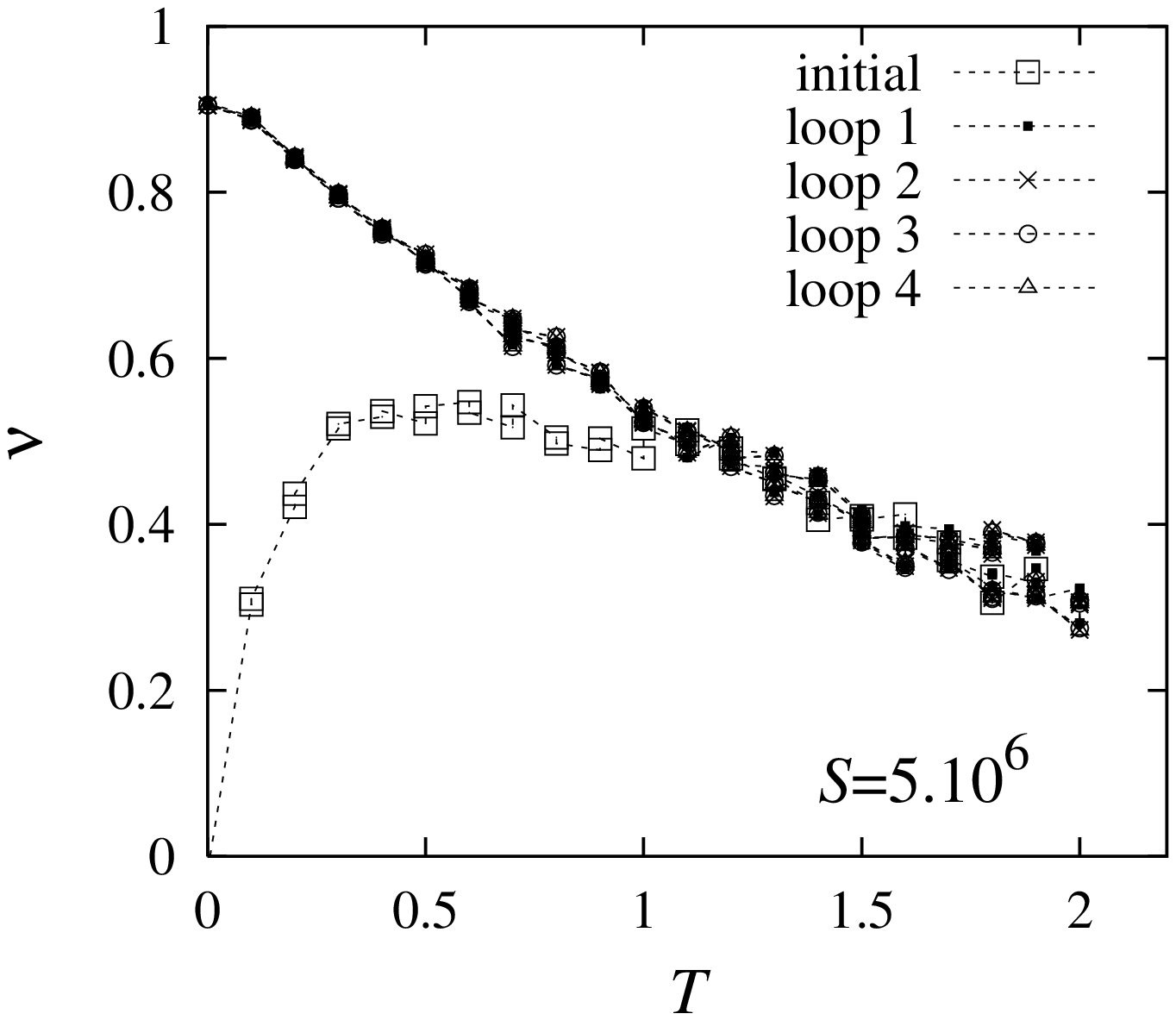,height=6.0cm,angle=-90}
\end{center}
\caption{Rescaled density $\nu$ plotted as a function of
the temperature $T$ for a periodic temperature sequence.
The arrows indicate the direction of the loops.
}
\label{fig:cycle}
\end{figure}

In parallel to the tapping experiments by Nowak et al.~\cite{nowak97},
we present simulations of our model using a periodic time series for
the temperature.  The temperature is kept constant for $S$ steps and then
increased by $\Delta T=0.1$, where it is kept again constant for $S$
steps. $T$ is initially zero and then raised up to $T=2$. From this state,
$T$ is decreased to zero and the loop is repeated seven times.
In Fig.\ \ref{fig:cycle}, the simulation results are displayed for
different $S$ as given in the inset.  In the initial branch with 
increasing $T$, the density increases and slowly decreases for large
$T$.  This branch is irreversible, but the periodic loops show almost
reversible behavior. For very short loops (small $S$), the density
continuously increases, for longer loops the behavior of the system
is reversible. Note that the system also shows hysteretic behavior,
the density at decreasing $T$ is below the density at increasing $T$.

In summary, we presented a very simple model for the dynamics of
the compaction of granular media due to an external agitation.
Our model is not as detailed as others (parking lot or frustrated
lattice-gas models), but it is extremely simple and still shows
qualitative agreement with two different types of experiment. 
Its simplicity allows for future analytical treatments. However, such 
a simple model arises, besides many others, two major questions: 
(i) Is it possible to link the 
real configuration phase space with the an energy landscape? (ii) 
Is it correct to do the analogy between the experimental external 
excitations like shear or tapping  with a temperature?


\section*{Acknowledgements}
S. L. acknowledges the support of the I.U.S.T.I., Marseille and
thanks for the hospitality. Furthermore, helpful discussions with 
A. Blumen and I. Sokolov are appreciated.

\end{document}